\documentclass[12pt]{article}
\usepackage{graphicx,tabularx,epsfig}
\usepackage{color}
\usepackage{enumitem}
\usepackage{caption}
\usepackage{subcaption}
\usepackage{amsmath}
\usepackage{amssymb}
\usepackage{amsthm}
\usepackage{physics}
\usepackage{dsfont, mathtools}
\usepackage{psvectorian}
\usepackage{blkarray}
\usepackage{bm}
\usepackage{url}
\usepackage{setspace}

\usepackage{xcolor}
\colorlet{linkequation}{blue}

\usepackage[colorlinks = true,
            linkcolor = blue,
            urlcolor  = blue,
            citecolor = blue,
            anchorcolor = blue]{hyperref}

\usepackage{floatrow}

\newlength{\abstractwidth}
\renewcommand{\thefootnote}{\fnsymbol{footnote}}
\renewcommand{\thanks}[1]{\footnote{#1}} 
\newcommand{\starttext}{
\setcounter{footnote}{0}
\renewcommand{\thefootnote}{\arabic{footnote}}}

\makeatletter
\g@addto@macro\normalsize{%
  \setlength\abovedisplayskip{10pt}
  \setlength\belowdisplayskip{20pt}
  \setlength\abovedisplayshortskip{10pt}
  \setlength\belowdisplayshortskip{20pt}
}
\makeatother

\usepackage{color}
\DeclareMathOperator{\arccosh}{arccosh}

\renewcommand{\title}[1]{\vbox{\center\LARGE{#1}}\vspace{5mm}}
\renewcommand{\author}[1]{\vbox{\center#1}\vspace{5mm}}

\numberwithin{equation}{section}

\begin{document}

\singlespacing

\begin{center}

{\Large \bf {New bulk cone singularities in Vaidya-like spacetimes from large $c$
conformal blocks}}

\bigskip \noindent
\bigskip
\bigskip

 {Henry Leung$^a$}

\bigskip
\bigskip

    {\it $^a$  Department of Physics, University of California, Santa Barbara, CA 93106}

\bigskip
\bigskip
    

\bigskip
\bigskip
\bigskip

\end{center}

\begin{abstract}
Bulk cone singularities are singularities in boundary two-point functions at points separated by a null geodesic in the bulk, but not in the boundary. In this work, we describe a new type of bulk cone singularities in a family of Vaidya-like spacetimes that are labeled by the radius $r_+$ of the resulting black hole. We find a sharp transition in the causal structure within this family of spacetimes at $r_+=l$, the AdS length. In particular, there are bulk cone singularities that do not exist in the $r_+>l$ case, but appear for $r_+<l$. In the case of $r_+<l$ where such singularities exist, we are able to reproduce the singularities in a CFT$_2$ calculation using large $c$ conformal blocks.

\medskip
\noindent
\end{abstract}

\newpage

\starttext \baselineskip=17.63pt \setcounter{footnote}{0}

{\hypersetup{hidelinks}
\tableofcontents
}


\starttext \baselineskip=17.63pt \setcounter{footnote}{0}


\section{Introduction}
\label{sec:intro}
A hallmark of a large $N$ holographic theory is the emergence of a bulk causal structure. This leads to striking features in boundary correlators. In any QFT, correlator with insertions that are null separated exhibits singularities. However, in holographic CFTs, in a state with a semiclassical bulk dual, there are additional bulk null geodesics, which may enlarge the set of null separated points on the boundary. This happens in particular when the bulk is nontrivial and focusing causes a bulk null geodesic to reach the same boundary spatial point at a different time compared to a boundary null geodesic, as opposed to the situation in global AdS where bulk and boundary null geodesics reach the same boundary point at the same time. The presence of additional singularities does not contradict the boundary causal structure, since the Gao-Wald theorem guarantees that the bulk cone singularities never appear before the light cone singularities \cite{Gao:2000ga}. Due to these bulk null geodesics, boundary correlators develop new singularities \cite{Hubeny:2006yu,Gary:2009ae,Maldacena:2015iua}. There are also cases where boundary points are separated via a nearly null geodesic, often present due to black hole singularities. While these typically do not lead to obvious singularities in the correlator, they do leave behind signatures that are characteristic of an emergent bulk \cite{Fidkowski:2003nf,Festuccia:2005pi,Amado:2008hw,Horowitz:2023ury}. Although these new features are natural from a bulk point of view, they remain mysterious in the CFT, especially since there are various arguments that indicate that they cannot appear in a CFT at finite $N$ and finite 't Hooft coupling \cite{Hubeny:2006yu,Maldacena:2015iua,Dodelson:2020lal}.

\begin{figure}[H] 
 \begin{center}                      
      \includegraphics[width=3in]{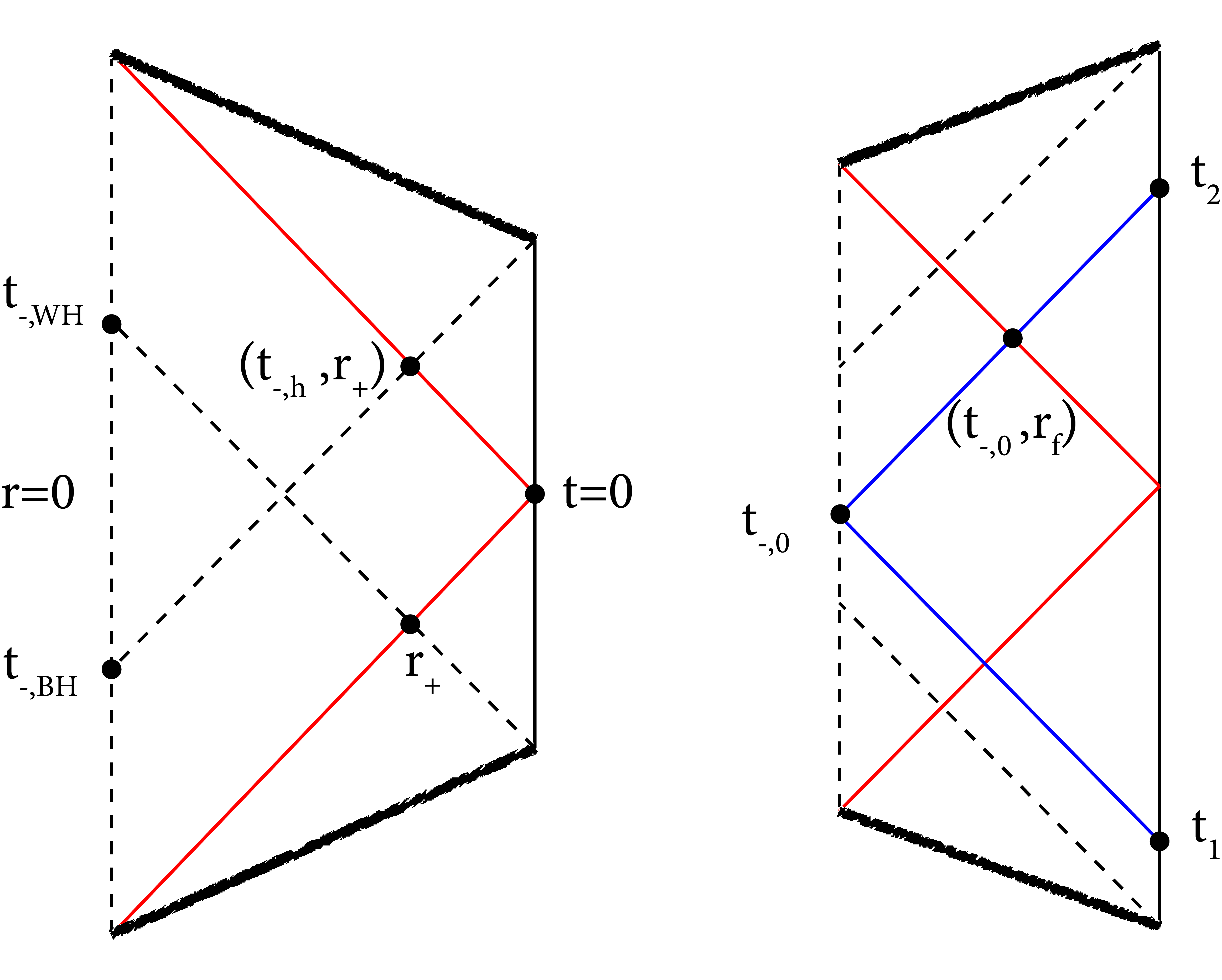}
      \caption{The Penrose diagram for the family of geometries described in this section. The red line represents the shockwave, which comes out of a white hole, reaches the boundary at $t=0$ and collapses into a black hole. The geometry is global AdS inside the shock wave and is Schwarzschild or BTZ outside the shock wave. The vertical dotted line represents the origin $r=0$ and the diagonal dotted lines represents the black hole and white hole horizons. On the left, we have the case of $r_+>l$, so that the black hole and white hole horizons overlap. On the right, we have $r_+<l$ where there are Cauchy slices that do not intersect either horizons.}
  \label{fig:spacetime}
  \end{center}
\end{figure}

In this paper, we approach this problem from a CFT point of view. We focus on the simplest case of such new singularities -- the bulk cone singularity, which arises when a boundary two-point function has insertions that are separated by a bulk null geodesic. The most well-studied bulk cone singularities are in static black hole backgrounds \cite{Hubeny:2006yu,Dodelson:2023nnr}, but the phenomenon shows up more generally. In this work, we study a particular bulk geometry, given by a time-reversal symmetric, spherical, null shock wave acting on global AdS, i.e. the null shell emerges out of a white hole, bounces at the boundary, and collapses into a black hole. This family of simple geometries has a feature that gives rise to an interesting bulk cone structure. If the shock wave creates a black hole that has $r_+>l$, the caustic of the black hole horizon is inside of the white hole; whereas if the black hole has $r_+<l$, the black hole and the white hole do not overlap (see Fig. \ref{fig:spacetime}). This means that in the case of the bigger black hole, a radial null geodesic sent from the boundary always enters the black hole, but a radial null geodesic in the geometry with $r_+<l$ can escape to the boundary given that it is sent in early enough. So within this family of geometry, there is a sharp transition at $r_+=l$ that determines whether a boundary two-point function has the bulk cone singularities associated with the radial null geodesic mentioned above. Notice the difference between the bulk cone singularities studied in this work and those in \cite{Dodelson:2023nnr}, which results from null geodesics that carry angular momentum. In a static black hole background, it is not possible to have radial geodesics that give rise to bulk cone singularities. This is also to say that the bulk cone singularities due to null geodesics that carry angular momentum would still be present in the family of geometries that we consider, both below and above the threshold $r_+=l$, and the transition that we find is related to the presence/absence of the class of radial bulk cone singularities.

The spacetime has a natural description on the CFT as a state given by a heavy operator $V$ acting on the vacuum at $t=0$. A two-point function in the geometry is then given by the following correlator in the CFT
\begin{equation}\label{eq:corr}
    \langle 0\vert V^{\dagger}(t=0)O(t_1,0)O(t_2,\theta)V(t=0)\vert 0\rangle\,,
\end{equation}
where $O$ is a probe operator. While the family of geometries mentioned exists in AdS$_{d+1}$/CFT$_{d}$ for any $d\geq 2$, we mostly restrict to the case of $d=2$. Our main result is a calculation of the above correlator in large $c$ CFT$_2$ that reproduces the bulk result, i.e. the presence/absence of bulk cone singularities below/above the transition between the family of geometries at $r_+=l$, represented by the conformal dimension of $V$.

We calculate the correlator \eqref{eq:corr} following the methods in \cite{Anous:2016kss,Anous:2017tza}. A CFT correlator can be expanded in a fixed channel as a sum of conformal blocks over all primaries, but \cite{Anous:2017tza} suggests that, under mild assumptions, this is well-approximated by a sum of the Virasoro identity blocks over all channels. Since we are interested in the large $c$ limit, the identity blocks can be calculated by the monodromy method developed in \cite{Zamolodchikov:1987avt,Hartman:2013mia}. By examining the sum over the identity blocks, we are able to identify a divergence in the CFT correlator at a time that matches with the time of the bulk cone singularity from the bulk calculation.

The paper is organized as follows. In Section \ref{sec:bulk}, we introduce the family of geometries and the bulk cone singularities of interest. We then calculate the relevant CFT correlator that represents the bulk effect in section \ref{sec:cft}. Next, in Section \ref{sec:analysis}, we analyze the singularities of the correlator and match them with the bulk picture. We also give a saddle point analysis of the correlator at large $\Delta$. We conclude with a discussion in Section \ref{sec:disc}.

\section{Bulk spacetime}
\label{sec:bulk}

As mentioned in the introduction, we are interested in a family of spacetimes corresponding to a time-reversal symmetric, spherical, null shock wave acting on global AdS. We denote the quantities inside the shock wave with subscript ``$-$'' and outside the shock wave with subscript ``$+$''. Time coordinates with neither subscripts are reserved for boundary times. The metric is Schwarzschild/BTZ on the outside, and global AdS on the inside
\begin{equation}
    ds_{\pm}^2=-f_{\pm} dt^2_{\pm}+f_{\pm}^{-1} dr^2+r^2 d\Omega_{d-1}^2\,,
\end{equation}
where $f_-=1+r^2/l^2$, and
\begin{equation}
    f_+=\begin{cases}
        \frac{r^2}{l^2}-\frac{\mu}{r^{d-2}}+1\,,\quad d>2\\
        \frac{r^2-r_+^2}{l^2}\,, \quad d=2\,,
    \end{cases}
\end{equation}
and $\mu$ is proportional to the mass of the black hole. In either case, we will use $r_+$ to denote the horizon, the outermost root of $f_+$. The inside and outside geometries are glued together along the hypersurface ruled by radial null geodesics using null junction conditions, which implies in particular that the radial coordinate is continuous across the junction. Since the shock wave is time-symmetric, it has both future and past portions. We will choose the time that the shock wave hits the boundary to be $t_\pm=0$.

\begin{figure}[H] 
 \begin{center}                      
      \includegraphics[width=4in]{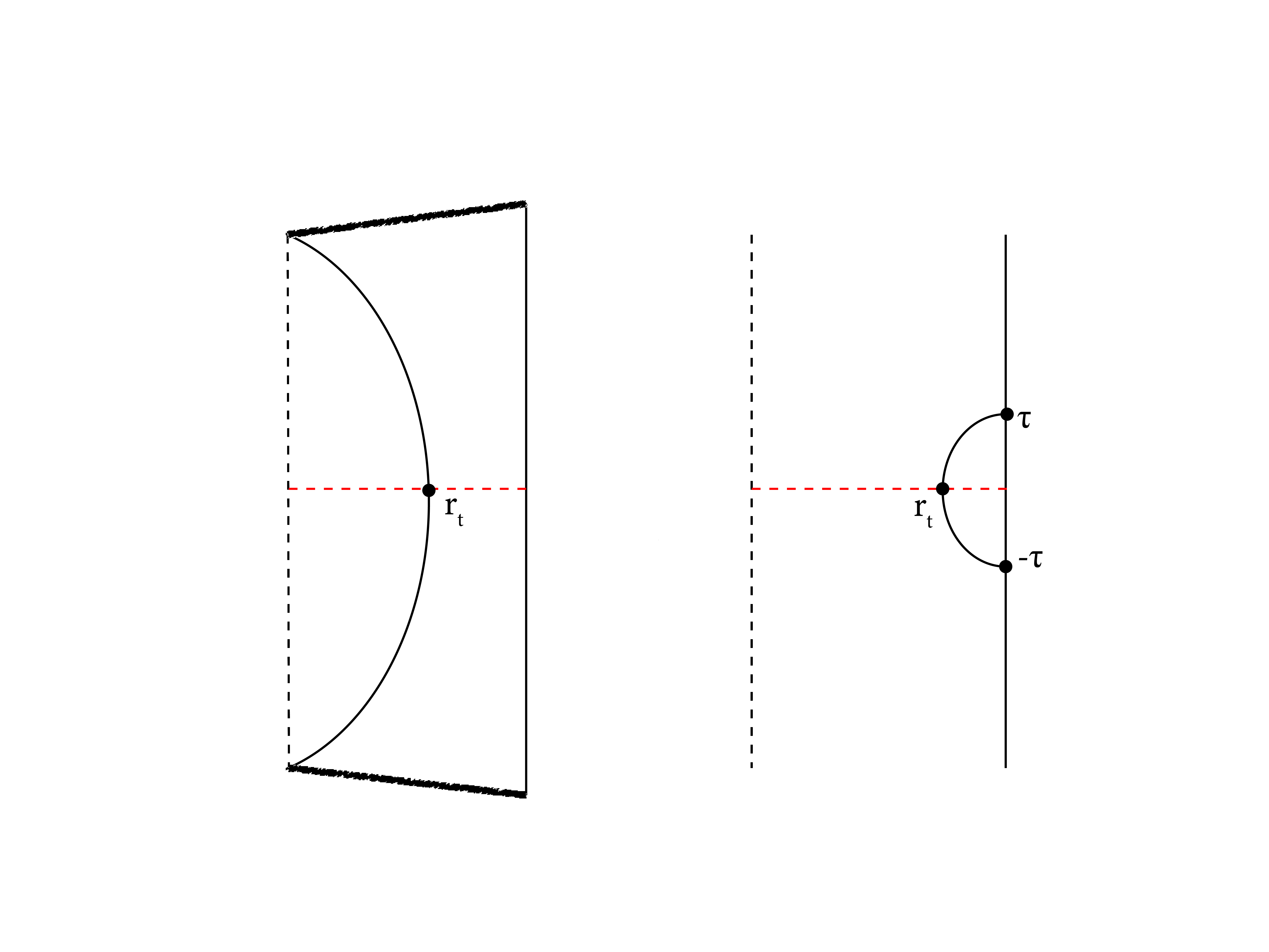}
      \caption{Left: the null shock wave geometry can be obtained as limit of timelike shells with increasing turning points $r_t$. Right: Wick-rotation gives a Euclidean geometry with a shell that reaches the boundary at $\pm \tau$.}
  \label{fig:timelike-euclidean}
  \end{center}
\end{figure}

The geometry with the null shock wave can be obtained as a limit of the geometry glued together with a timelike dust shell. The timelike shell is parameterized by its proper mass and has a turning point on the time-symmetric slice at finite $r=r_t$. The null shock wave limit is then obtained by a vanishing proper mass, which corresponds to the turning point approaching the boundary. Consider the initial data on the time-symmetric slice for the family of geometries with the timelike shell. The metric can be wick-rotated to Euclidean signature simply by taking $t_\pm\to -i\tau_\pm$. As opposed to the Lorentzian timelike shell, the trajectory for the analytically-continued Euclidean shell reaches the boundary as shown in Fig. \ref{fig:timelike-euclidean}. The limit in which the shell becomes null upon wick-rotation is given by taking the Euclidean times at which the shell hits the boundary to coincide. This will become important in the next section since we will be doing the CFT calculation in Euclidean signature.

It is useful to introduce Kruskal coordinates on the Schwarzschild/BTZ side. The tortoise coordinate is defined by
\begin{equation}
    r_*(r)=-\int_r^\infty \frac{dr'}{f_+(r')}\,.
\end{equation}
The Kruskal coordinates are defined as
\begin{equation}
    U=- e^{\frac{2\pi}{\beta} (r_*(r)-t_+)}\,,\quad V=e^{\frac{2\pi}{\beta}(r_*(r)+t_+)}\,.
\end{equation}
The null shockwave can then be easily described as the future piece $V=1$ and the past piece $U=-1$.

Note that the mass/radius of the black hole is arbitrary.\footnote{For $d=2$, we require the geometry outside to be above the BTZ threshold.} We now show that there is a sharp distinction between the geometries where the black hole has $r_+>l$ and those with $r_+<l$. On the outside, the black hole and white hole horizons are given by $r=r_+$, but as one moves along these hypersurfaces across the shock wave into global AdS, the area of these horizons shrink and eventually the congruence of geodesics that rule the horizon intersects and forms a caustic. The relative times of the caustics for the black hole and white hole horizons determine whether the they overlap. Thus, we want to calculate the (inside) time $t_{-,BH}$ at which the black hole horizon forms a caustic at the origin of global AdS. On the side of global AdS, the null geodesics along the horizon reaches $r_+$ at time $t_{-,h}$ given by
\begin{equation}\label{eq:t-origin-hor}
    t_{-,h}-t_{-,BH}=\int_0^{r_+} \frac{dr}{f_-(r)}=l\tan^{-1}\frac{r_h}{l}\,.
\end{equation}
This time is also when the shock wave sent from the boundary at $t=0$ meets the horizon and is given by
\begin{equation}\label{eq:t-bdy-hor}
    t_{-,h}=-\int_{\infty}^{r_+} \frac{dr}{f_-(r)}=l\qty(\frac{\pi}{2}-\tan^{-1}\frac{r_h}{l})\,.
\end{equation}
Together, they gives the time of the caustic as 
\begin{equation}
    t_{-,BH}=l\qty(\frac{\pi}{2}-2\tan^{-1}\frac{r_h}{l})\,.
\end{equation}
Note that $t_{-,BH}>0$ when $r_+<l$ and $t_{-,BH}<0$ when $r_+>l$. Since the setup is time-symmetric, we get the time of the white hole horizon caustic as $t_{-,WH}=-t_{-,BH}$, which means that the black hole and white hole horizons overlap if and only if $r_+>l$.

This simple fact has interesting physical consequences. It implies that when $r_+<l$, it is possible for a radial null geodesic to start at a boundary point and end up at an antipodal boundary point given that it is sent early enough, whereas no such radial null geodesic is possible for $r_+>l$ since they must all enter the black hole. This gives rise to bulk cone singularities that are present if and only if $r_+<l$. While we focus on the bulk cone singularities in this paper, we point out in the discussion \ref{sec:disc} another interpretation of the transition at $r_+=l$ in terms of the single trace algebra in this state, using the language of \cite{Leutheusser:2022bgi}.

Now we want to obtain more concrete information about the bulk cone singularity in the case of $r_+<l$. More precisely, suppose we send a radial null geodesic before the shock wave from the boundary, at $t_+=t_1<0$, we want to calculate at what time $t_+=t_2>0$ does the geodesic reach the boundary at an antipodal point, if it does at all. Clearly, a prerequisite for such a geodesic to exist is that the time $t_{-,0}$, at which the geodesic arrives at the origin of AdS, must be between the caustics of the black and white holes horizons, i.e. $t_{-,WH}<t_{-,0}<t_{-,BH}$. We first find $t_2$ in terms of $t_{-,0}$. Let the inside time and radius at which the geodesic hits the future piece of the shock wave be $t_{-,f}$ and $r_f$. Following the same calculations in \eqref{eq:t-bdy-hor} and \eqref{eq:t-origin-hor}, we can get $t_{-,0}$ in terms of $r_f$, which can be inverted to give
\begin{equation}
    r_{f}=l\tan\qty(\frac{\pi}{4}-\frac{t_{-,0}}{2l})
\end{equation}
As mentioned, the radial coordinate $r$ is matched inside and outside the shell by the junction conditions. So using Kruskal coordinates and the fact that we are on the piece of the shock wave given by $V=1$, we have at the intersection
\begin{equation}
    U=UV=-e^{\frac{4\pi}{\beta} r_*(r_f)}\,.
\end{equation}
Once this outgoing geodesic crosses the shock wave, it stays at constant $U$. So at the boundary, we have
\begin{equation}
    -e^{\frac{4\pi}{\beta} r_*(r_f)}=U=-e^{\frac{2\pi}{\beta} (r_*-t_2)}|_{r_*=0}
\end{equation}
which gives
\begin{equation}
    t_2=-2r_*\qty(l\tan\qty(\frac{\pi}{4}-\frac{t_{-,0}}{2l}))\,.
\end{equation}
A similar calculation to relate $t_1$ to $t_{-,0}$ gives
\begin{equation}
    t_1=2r_*\qty(l\tan\qty(\frac{\pi}{4}+\frac{t_{-,0}}{2l}))\,.
\end{equation}
We can eliminate $t_{-,0}$ to get
\begin{equation}
    t_2(t_1)=-2r_*\qty(
    \frac{l^2}{r\qty(\frac{t_1}{2})}
    )\,,
\end{equation}
where $r(\cdot)$ denotes the inverse function of $r_*(r)$. As $t_1\to-\infty$, we get $t_2(-\infty)=-2r_*(l^2/r_+)$, which is the earliest time a bulk cone singularity can appear. If we write $l=\gamma r_+$, we get $t_2(-\infty)=-2r_*(\gamma^2 r_+)$, which is only real for $\gamma<1$ and diverges as $\gamma$ approaches 1 from below, in agreement with the previous analysis that the bulk cone singularity is only present for $r_+<l$. Finally, we state explicitly the time for the bulk cone singularity in the case of BTZ, which we will reproduce in a CFT calculation in the rest of this paper
\begin{equation}\label{eq:BC-time}
    \frac{t_{2}(t_1)}{l}=\frac{l}{r_+}\log\qty(\frac{\tanh\frac{r_+ t_{1}}{2l^2}-\frac{r_+^2}{l^2}}{\tanh\frac{r_+ t_{1}}{2l^2}+\frac{r_+^2}{l^2}})\,.
\end{equation}

\section{The CFT calculation}\label{sec:cft}
\subsection{The setup}
In this section, we will describe the CFT correlator that is dual to the two-point function in the family of geometries that we described. This section closely follows \cite{Anous:2016kss}, which has a very similar setup but with different operator insertions. It is nonetheless useful to review the setup and comment on the difference. We will restrict to the case of $d=2$ in which the CFT calculation is done. The geometry mentioned above is dual to the vacuum state acted on by a spatially homogeneous heavy operator $V$ at Lorentzian time $t=0$
\begin{equation}
    \vert \chi\rangle=V(t=0)\vert 0\rangle\,.
\end{equation}
The boundary two-point Wightman function of probe operators at Lorentzian times in the geometry correspond to the Wightman function
\begin{equation}\label{eq:gen-corr}
    G(t_1,t_2)=\langle \chi\vert O_2(t_2)O_1(t_1)\vert \chi\rangle=\langle 0\vert V(t=0)^\dagger O_2(t_2)O_1(t_1) V(t=0)\vert 0\rangle\,.
\end{equation}
Here we have suppressed spatial dependence for simplicity. Note that the probe operators always see the effect of the shock wave operator $V$ regardless of whether they are inserted at a time before or after the shock wave, which is reflective of the fact that the shock wave in our geometry is time-symmetric, as opposed to the more traditional Vaidya spacetimes with a shock wave only in the future, which corresponds to a different operator ordering studied in \cite{Anous:2016kss}.\footnote{The operator ordering in the Vaidya shock wave spacetime is
\begin{align*}
    \langle 0\vert V^\dagger O_2(t_2) VO_1(t_1)\vert 0\rangle\,,\quad t_2>0&\,,t_1<0\,,\\
    \langle 0\vert O_2(t_2)V^\dagger VO_1(t_1)\vert 0\rangle\,,\quad t_2<0&\,,t_1<0\,,\\
    \langle 0\vert V^\dagger O_2(t_2) O_1(t_1) V^\dagger\vert 0\rangle\,,\quad t_2>0&\,,t_1>0\,,
\end{align*}
whereas the operator ordering is always fixed to be \eqref{eq:gen-corr} in our case regardless of $t_1,t_2$, but we will only consider $t_1<0$ and $t_2>0$.}

\begin{figure}[H] 
 \begin{center}                      
      \includegraphics[width=2.5in]{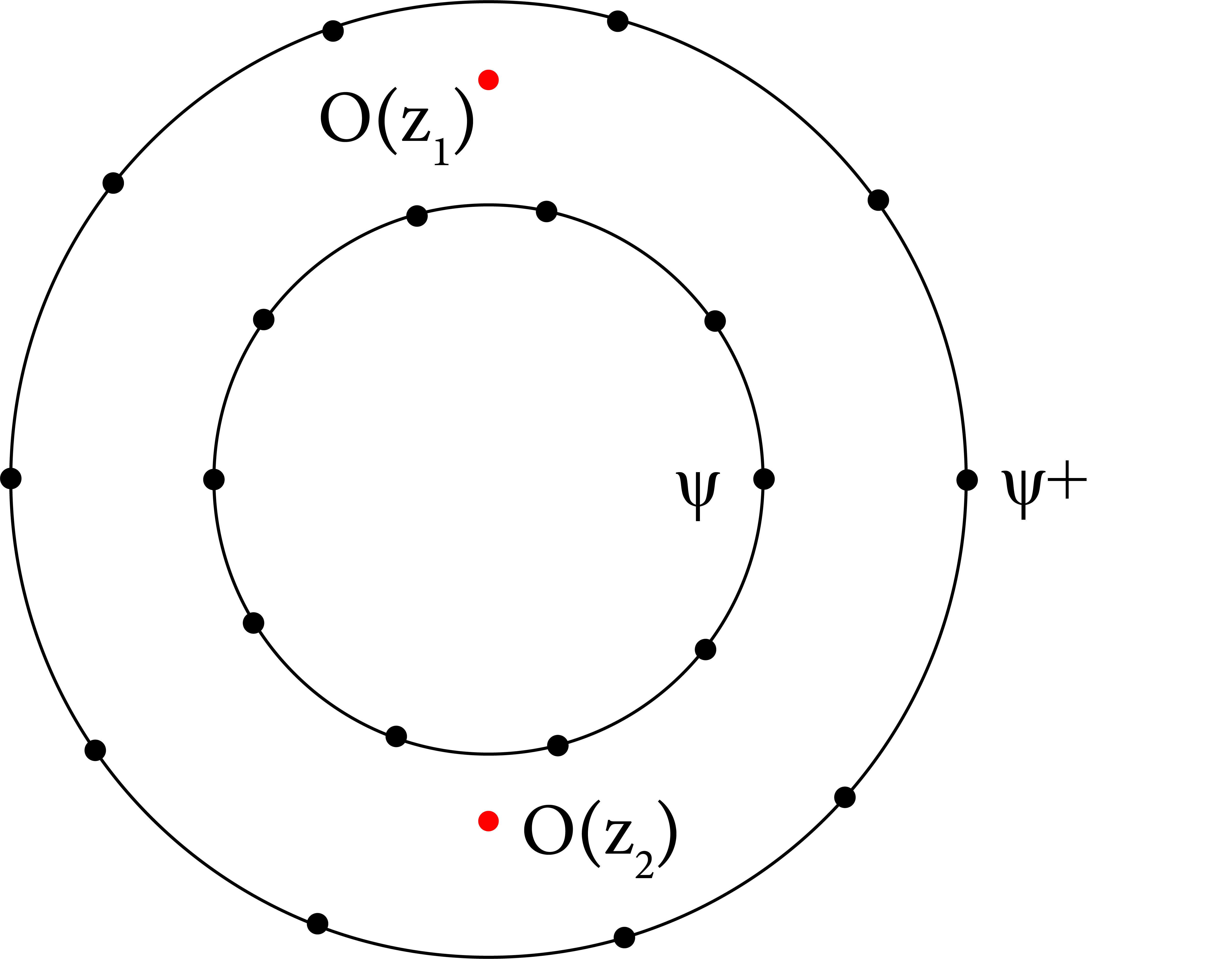}
      \caption{The operator insertions in \eqref{eq:many-pt-func}. The two rings of operators create the state and the red points are the probe operators. These are operators in the CFT on the plane. By going to cylindrical coordinates, we obtain the boundary of the figure on the right in Fig. \ref{fig:timelike-euclidean}.}
  \label{fig:CFT-insertions}
  \end{center}
\end{figure}

While the bulk cone singularities that we would like to study are a Lorentzian effect, it is easier to carry out the CFT calculation in Euclidean signature and then analytically continue to real time. As mentioned in the previous section, the null shock wave can be obtained as a limit of timelike shells. Such geometries can be wick-rotated to a Euclidean spacetime at the time-symmetric slice. The Euclidean shell reaches the boundary at $\pm \tau$ and the limit $\tau\to 0$ gives the null shock wave in the Lorentzian spacetime. The Euclidean times $\pm \tau$ thus represent two operator insertions on the vacuum state in the CFT on the cylinder $S^{d-1}\times \mathbb{R}$, corresponding to $V$ and $V^\dagger$ (regularized by Euclidean time $\tau$). To be more precise, we consider the operators in radial quantization and the two timeslices $\pm \tau$ on the cylinder become concentric spheres around the origin at radii $e^{\pm \tau}$, which reflects the fact that these are conjugate operators. We model the spatially homogeneous operators $V$ as a product of local operators, each representing a dust operator that make up the null shell in the bulk (see Fig. \ref{fig:CFT-insertions}). The $\tau$-regularized state is then
\begin{equation}
    V_{-\tau}\vert 0\rangle=\lim_{n\to \infty} \frac{1}{N_n} \prod_{k=1}^n \psi(e_k,\bar{e}_k)\vert 0\rangle\,,\quad e_i=e^{-\tau+2\pi i\frac{k-1}{n}}\,,
\end{equation}
where $N_n$ is a normalization factor. Using the fact that the conjugate operator of any operator $Q$ is defined as $Q(z,\bar{z})^\dagger=z^{-2\bar{h}_Q}\bar{z}^{-2h_Q}Q(\bar{z}^{-1},z^{-1})$ in radial quantization, we find that \eqref{eq:gen-corr} is given by an analytic continuation of the vacuum correlator
\begin{equation}\label{eq:many-pt-func}
    G(z_1,\bar{z}_1;z_2,\bar{z}_2)=\lim_{\tau\to 0}\lim_{n\to\infty} \frac{1}{\abs{N_n}^2}\prod_{i,k}e_i^{-2\bar{h}_\psi}\bar{e}_i^{-2h_\psi}\langle\psi(\bar{e}_i^{-1},e_i^{-1}) O_2(z_2,\bar{z}_2)O_1(z_1,\bar{z}_1)\psi(e_k,\bar{e}_k)\rangle\,.
\end{equation}
It should be noted that Euclidean correlators are always time-ordered, so in order to get the operator ordering in \eqref{eq:gen-corr}, we must have $\tau>\tau_1>\tau_2>-\tau$, where $z_i=e^{\tau_i+i \theta_i}$. Using rotation symmetry, we can label the angles as $z_1=e^{\tau_1+i \theta/2}$ and $z_2=e^{\tau_2-i \theta/2}$. The Lorentzian correlator is then obtained by the analytic continuation $\tau_i\to \tau_i+i t_i$. The case that we are interested in involves two operator insertions at antipodal points in the state $\vert \chi\rangle$, so eventually we will take $\theta=\pi$.

\subsection{Correlator via Virasoro identity block}
In this subsection, we will calculate the correlator \eqref{eq:many-pt-func} using the methods in \cite{Anous:2016kss,Anous:2017tza}. It has been shown in \cite{Hartman:2013mia} that Euclidean correlators are dominated by Virasoro identity blocks. We denote the holomorphic and anti-holomorphic parts of the vacuum block by $\mathcal{F}_0$ and $\overline{\mathcal{F}}_0$. By crossing symmetry, the vacuum block can only dominate in a particular channel $\Gamma$, i.e. the one in which the vacuum block contribution is largest, thus correlators can be approximated by
\begin{equation}\label{eq:max-channel}
    G(z_1,\bar{z}_1;z_2,\bar{z}_2)\approx\max_{\Gamma} \abs{\mathcal{F}_0^\Gamma}^2\,.
\end{equation}
However, this is no longer true in Lorentzian signature, where the correlator is not dominated by a single channel. Instead, following the general idea of ``black hole farey tail''\cite{Dijkgraaf:2000fq,Maloney:2007ud,Maloney:2016kee,Fitzpatrick:2011ia}, the authors in \cite{Anous:2017tza} suggest to approximate the correlator as a sum of the vacuum blocks over various channels $\Gamma$
\begin{equation}\label{eq:sum-channels}
    G(z_1,\bar{z}_1;z_2,\bar{z}_2)\approx\sum_{\Gamma} \abs{\mathcal{F}_0^\Gamma}^2\,,
\end{equation}
which would reduce to \eqref{eq:max-channel} when one channel dominates. This is quite different from the more conventional sum over primaries in CFT
\begin{equation}\label{eq:sum-primaries}
    G(z_1,\bar{z}_1;z_2,\bar{z}_2)=\sum_{\textnormal{primaries}~O_p} \abs{\mathcal{F}_{p}^\Gamma}^2
\end{equation}
but \cite{Anous:2017tza} argues that \eqref{eq:sum-channels} is a good approximation to \eqref{eq:sum-primaries} in holographic theories based on mild assumptions. In this paper, we will follow their approach and find good agreement between the results and the expectation from gravity, which further supports their proposal.

\begin{figure}[H] 
 \begin{center}                      
      \includegraphics[width=5in]{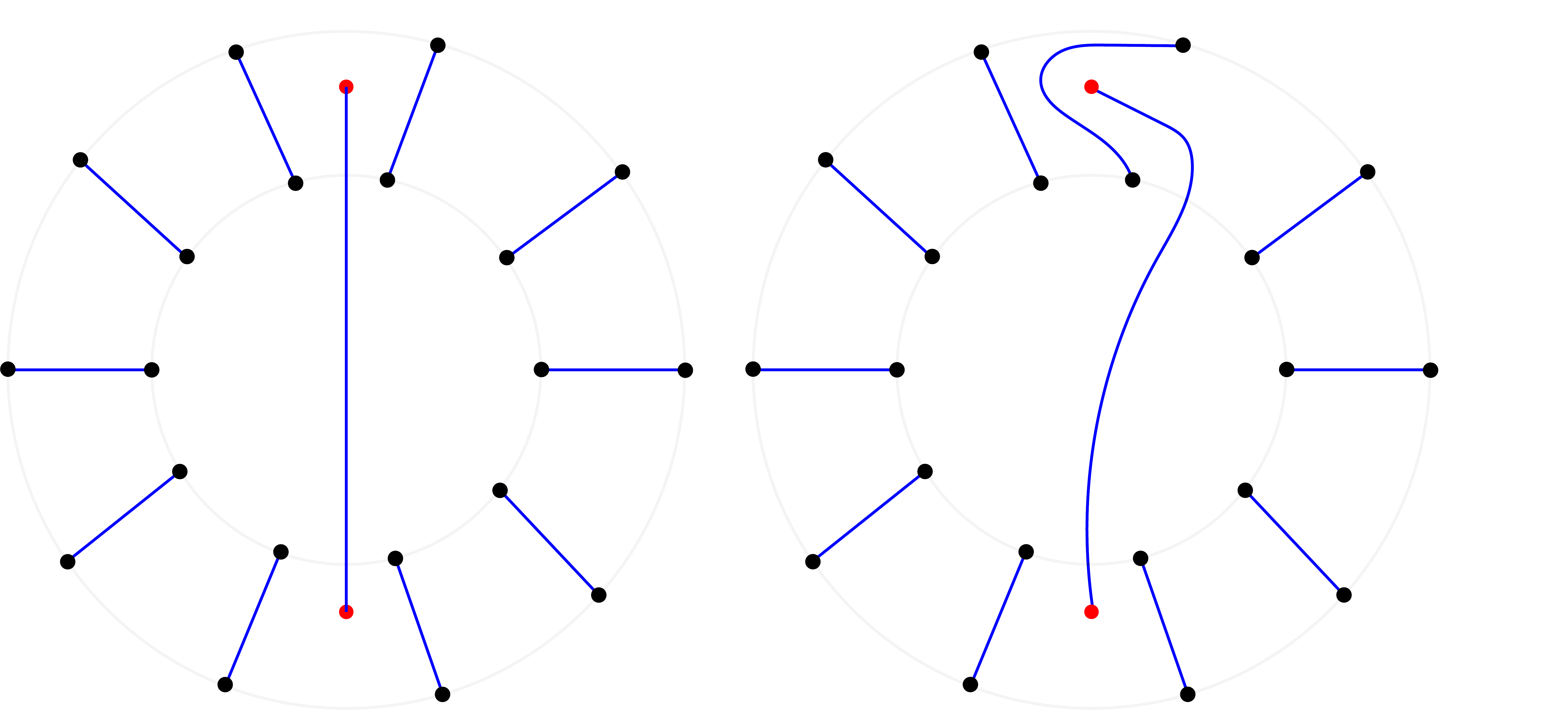}
      \caption{The blue lines here show the paths by which operators are contracted with each other. We show two different channels where the pairing of the operators are the same, but the paths are different. In the monodromy calculation, we require the monodromy of the solution to be trivial for non-intersecting cycles around each blue line.}
  \label{fig:channels}
  \end{center}
\end{figure}

Therefore, it remains to find the identity blocks in channels where it might dominate and sum over all such contributions. By dimensional arguments, these are channels where each dust operator is contracted with its complex conjugate and the probe operators are contracted with each other. As explained in \cite{Anous:2016kss}, this does not fully specify the channel, as one also needs to specify the paths by which the operators are brought together. Two such channels are shown in Fig. \ref{fig:channels}. The distinction between these channels will be more concrete in the following subsections, where we use the monodromy method to calculate the contribution from each channel.

\subsection{The monodromy method}
We now use the monodromy method to calculate the vacuum block contribution $G_\Gamma(z_1,\bar{z}_1;z_2,\bar{z}_2)$ to the two-point function in channel $\Gamma$. The monodromy method is used to calculate large $c$ conformal blocks and is developed in \cite{Zamolodchikov:1987avt} and reviewed in \cite{Harlow:2011ny,Fitzpatrick:2014vua}. Since our particular correlator corresponds to the setup in \cite{Anous:2016kss} with different operator insertions, our treatment here will be brief and the interested reader may consult \cite{Anous:2016kss} for further details.

We start with the equation
\begin{equation}\label{eq:mono}
    V''(z)+T_{cl}(z)V(z)=0
\end{equation}
where
\begin{equation}
    T_{cl}(z)=T_H(z)+T_L(z)\,.
\end{equation}
\begin{equation}
    T_{H}(z)=\sum_{k=1}^{2n} \qty{\frac{6h_\psi/c}{(z-z_k)^2}+\frac{c_k}{z-z_k}}\,.
\end{equation}
\begin{equation}
    T_{L}(z)=\sum_{m=1}^2 \qty{\frac{6h/c}{(z-z_m)^2}+\frac{c_m}{z-z_m}}\,.
\end{equation}
Here $h_\psi$ is the conformal weight of each dust operator $\psi_k$, $h$ is the conformal weight of each probe operator, and $c_k, c_m$ are accessory parameters that will be determined by the monodromy condition on the solution $\chi(z)$ specified by the channel $\Gamma$.

More concretely, the channel $\Gamma$ specifies the paths by which pairs of operators are contracted (see Fig. \ref{fig:channels}).\footnote{To fully specify a general channel, it is necessary to also specify the paths pairing up exchange operators, rather than just the external operators. However, here the $n$-point function is dominated by the identity block and it suffices to pair up the external operators.} For every path, we draw a closed, non-intersecting contour circling the operators involved in the contraction. The second order differential equation has two independent solutions $\chi_1(z)$ and $\chi_2(z)$, which undergoes a monodromy as they go around a loop $\gamma$ enclosing the singular points of $T_{cl}$
\begin{equation}
    \begin{pmatrix}
    \chi_1\\
    \chi_2
    \end{pmatrix}=M_{\gamma}\begin{pmatrix}
    \chi_1\\
    \chi_2
    \end{pmatrix}\,.
\end{equation}
We fix the accessory parameters by requiring that the monodromy to be trivial for every cycle given by the channel $\Gamma$.

The first term $T_H$ in $T_{cl}$ is related to the heavy operators, while the second $T_L$ is from the probe operators. $T_L$ will be treated perturbatively with $\epsilon=6h/c$ as the small parameter, i.e. we also have $c_i=O(\epsilon)$, whereas $T_H$ will be taken as $O(\epsilon^0)$ after accounting for other limits. In the continuum limit, $T_H$ becomes the following integral
\begin{equation}\label{eq:Th}
    \int_0^{2\pi}\frac{d\theta}{2\pi} \qty[\frac{6\hat{h}_\psi/c}{\qty(z-e^{\tau+i \theta})^2}-\frac{c_+(\theta)}{z-e^{\tau+i \theta}}+\frac{6\hat{h}_\psi/c}{z-e^{-\tau+i \theta}}-\frac{c_-(\theta)}{z-e^{-\tau+i \theta}}]\,,
\end{equation}
where the normalized weight $\hat{h}_\psi=n h_\psi$ is held fixed as $n\to \infty$. Rotational symmetry requires
\begin{equation}
    c_+(\theta)=-c_-(\theta)=K e^{-i\theta}\,,
\end{equation}
which together with the regularity condition $T_{cl}\sim z^{-4}$ gives the leading contribution of \eqref{eq:Th} in $\tau$ as
\begin{equation}\label{eq:Th-calc}
    \frac{K}{z^2}\Theta(\abs{z}-1+\tau)\Theta(1+\tau-\abs{z})+O(\tau)\,,
\end{equation}
where $K=\frac{6n h_\psi}{c\tau}$. As mentioned, we would like to solve the probe equation perturbatively in $O(\epsilon)$, so we must take $K$ to be $O(\epsilon^0)$ in the limits $n\to\infty$, $\tau\to 0$, $c\to \infty$. One can verify that this choice of accessory parameters is consistent with the monodromy associated with the channels specified in Fig. \ref{fig:channels}. \footnote{The two example channels both have the same set of contraction paths for the dust operators $\psi_k,\psi_k^\dagger$. The difference in monodromy arises only when the probe operators are taken into account at $O(\epsilon)$.} The dimensionless energy associated to this state on the cylinder is
\begin{equation}
    E=\frac{cK}{3}-\frac{c}{12}\,,
\end{equation}
which corresponds to a black hole of mass $m=E/l$. Using the fact that in BTZ $m=\rho^2/8G$ for $\rho\equiv r_+/l$, we find that the horizon radius is related to $K$ via
\begin{equation}
    \rho=\sqrt{4K-1}\,.
\end{equation}

The goal is to solve \eqref{eq:mono} perturbatively to $O(\epsilon)$. We denote the solution at $O(\epsilon^n)$ by $V^{(n)}$. We first use the result \eqref{eq:Th-calc} to solve for $V^{(0)}$. Outside the annulus, we have
\begin{equation}
    V^{(0)}_{o}=\begin{pmatrix}
        1 \\
        z
    \end{pmatrix}\,,
\end{equation}
and inside we have
\begin{equation}
    V^{(0)}_{i}=\begin{pmatrix}
        z^{\frac{1}{2}(1-i\rho)} \\
        z^{\frac{1}{2}(1+i\rho)}
    \end{pmatrix}\,.
\end{equation}
Across the two regions, the solution is continuous but has a discontinuous derivative. However, the discontinuity is subleading in $\tau$, so at leading order we can simply impose
\begin{equation}
    V^{(0)}_{i}(x_c)=J(x_c)V^{(0)}_{o}(x_c)\,,\quad V_{i}^{(0)'}(x_c)=J(x_c)V^{(0)'}_{o}(x_c)
\end{equation}
at the connection point $x_c$ in order to continue the solution into the other region. This then gives
\begin{equation}
    J(z)=\frac{1}{2}
    \begin{pmatrix}
         z^{\frac{1}{2}(1-i\rho)}(1+i \rho) && z^{-\frac{1}{2}(1+i\rho)}(1-i \rho)\\
         z^{\frac{1}{2}(1+i\rho)}(1-i \rho) && z^{-\frac{1}{2}(1-i\rho)}(1+i \rho)
    \end{pmatrix}
\end{equation}

We then add two probe operators $O$ to the mix. $T_L$ is
\begin{equation}
    \epsilon\qty(\frac{1}{(z-z_1)^2}+\frac{1}{(z-z_2)^2}-\frac{b_1}{z-z_1}-\frac{b_2}{z-z_1})\,,
\end{equation}
where $b_m=c c_m/6h$ are the rescaled accessory parameters.
Starting from the $O(1)$ solution at some reference point $z_0$, the $O(\epsilon)$ solutions in a neighborhood of $z_0$ inside and outside the annulus are
\begin{equation}
    V^{(1)}_{i}(z)=\mathcal{A}_{i}(\gamma_{z_0z})V^{(0)}_{i}(z)\,,\quad V^{(1)}_{o}(z)=\mathcal{A}_{o}(\gamma_{z_0z})V^{(0)}_{o}(z)\,,
\end{equation}
where
\begin{equation}
    \mathcal{A}_{i/o}(\gamma_{z_0z})=\qty(1+\epsilon\int_{\gamma_{z_0z}} F_{i/o})\,,
\end{equation}
here $\gamma_{z_0z}$ is a contour from $z_0$ to $z$, and
\begin{equation}
    F_{i}(z)=\frac{zT_L(z)}{i\rho}\qty(\begin{matrix}
        1 & -z^{-i\rho}\\
        z^{i\rho} & -1
    \end{matrix})\,,\quad F_{o}(z)=T_L(z)\qty(\begin{matrix}
        z & -1\\
        z^{2} & -z
    \end{matrix})\,.
\end{equation}
To cross from one region to another we again just need to multiply the solution at the connection point by $J(z_c)$ since it does not receive any correction at $O(\epsilon)$.

\begin{figure}[H] 
 \begin{center}                      
      \includegraphics[width=4in]{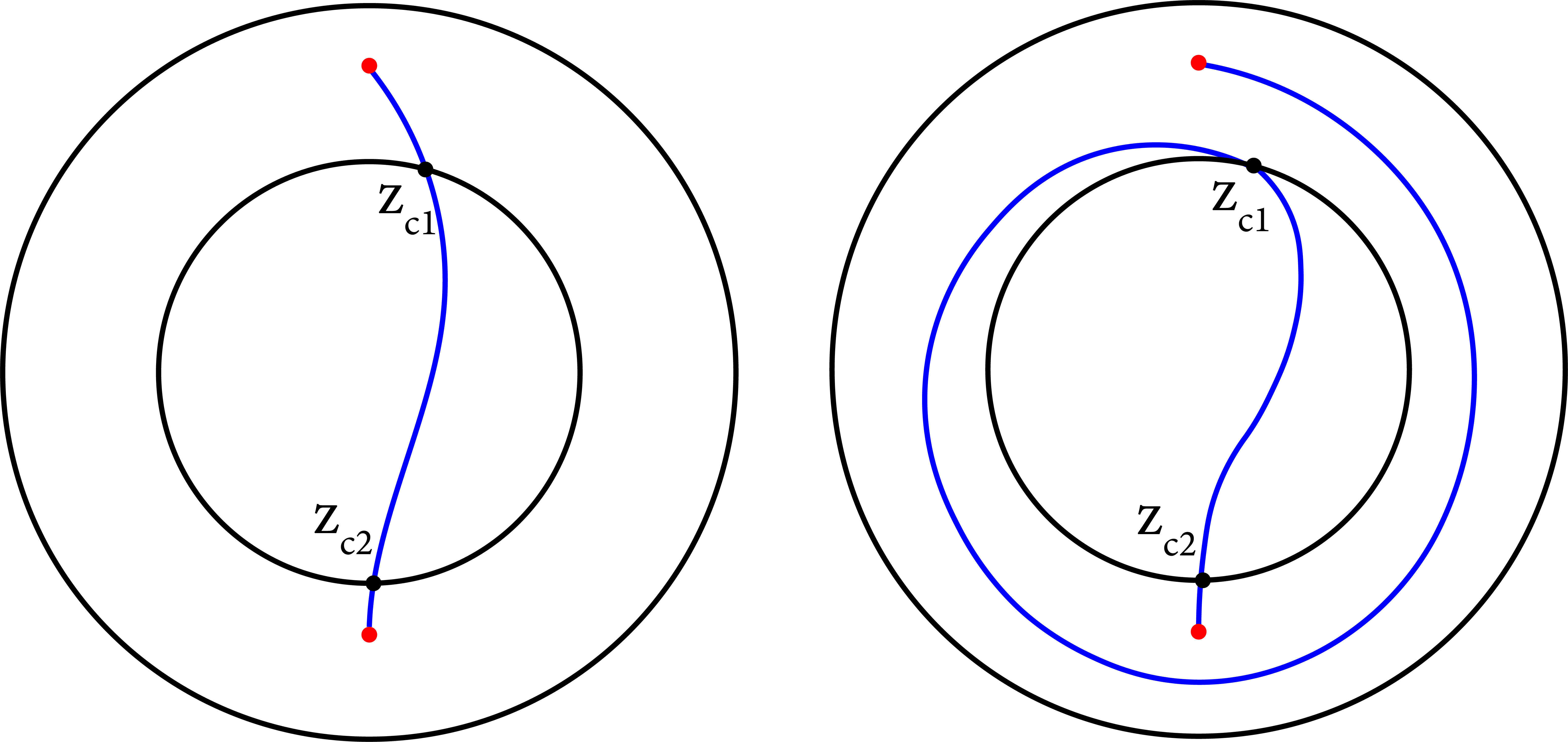}
      \caption{We denote the crossing points by $z_{c_1}$ and $z_{c_2}$. It is possible for contraction paths to have the same crossing points but distinct topology. On the right, the angle of $z_{c_1}$ is $2\pi$ less than the angle of $z_{c_1}$ in the left figure.}
  \label{fig:crossing-pts}
  \end{center}
\end{figure}

We can now consider the specific two-point function in Fig. \ref{fig:crossing-pts}, with the probes inserted at $z_1,z_2$ inside the annulus. The general contour enclosing the two points can start at $z_1$, intersect the inner shell at $z_{c_1}$ to move into the outside region, cross the shell again at $z_{c_2}$ to enclose $z_2$, and return to $z_1$ via the same connection points. We will solve the monodromy problem corresponding to the channel parametrized by each pair of connection points $z_{c_1},z_{c_2}$ and then integrate over all these channels. Note that although we parameterized the channels by assuming there are two crossings, the case where the contour does not intersect the shell at all is automatically included, since that is equivalent to the case where the crossing points degenerate $z_{c_1}=z_{c_2}$.

We can now calculate the monodromy for the contour in Fig. \ref{fig:crossing-pts}. Starting at some reference point $z_0$ in the outside region enclosed by the annulus, the solution transported around the contour is
\begin{equation}
    \mathcal{A}_{o}(\gamma_{z_0z_{c_1}}) J^{-1}(z_{c_1}) \mathcal{A}_{i}(\gamma_{z_{c_1}z_{c_1}}) J(z_{c_1}) \mathcal{A}_{o}(\gamma_{z_{c_1}z_{c_2}}) J^{-1}(z_{c_2}) \mathcal{A}_{i}(\gamma_{z_{c_2}z_{c_2}}) J(z_{c_2}) \mathcal{A}_{o}(\gamma_{z_{c_1}z_0})V^{(0)}_{o}(z_0)\,,
\end{equation}
where the contours $\gamma_{z_{c_1}z_{c_1}}$, $\gamma_{z_{c_2}z_{c_2}}$ return to the same connection point on the inner circle by enclosing $z_1$ and $z_2$ respectively. At $O(\epsilon)$, we have
\begin{equation}
    \qty(1+2\pi i\epsilon(J^{-1}(z_{c_1})\textnormal{Res}_{z_1}F_{i}\,J(z_{c_1})+J^{-1}(z_{c_2})\textnormal{Res}_{z_2}F_{i}\,J(z_{c_2})))V^{(0)}_{o}(z)+O(\epsilon^2)
\end{equation}
Trivial monodromy at $O(\epsilon)$ simply implies solving
\begin{equation}
    J^{-1}(z_{c_1})\textnormal{Res}_{z_1}F_{i}\,J(z_{c_1})+J^{-1}(z_{c_2})\textnormal{Res}_{z_2}F_{i}\,J(z_{c_2})=0
\end{equation}
for the rescaled accessory parameters $b_1,b_2$.

The holomorphic part of the vacuum block $\mathcal{F}_0=e^{-cf_0/6}$ is given by $b_1,b_2$ via
\begin{equation}\label{eq:f-from-accessory}
    \partial_k f_0(z_1, z_2)= \frac{6h}{c}b_k\,,
\end{equation}
supplemented by the boundary condition at coincident points
\begin{equation}
    f^{\Gamma(z_{c_1},z_{c_2})}_0(z_1, z_2)|_{z_1=z_{c_1},z_2=z_{c_2}}=\frac{12h}{c}\log(z_{c_1}-z_{c_2})\,\quad \textnormal{as}\quad z_{c_1}\to z_{c_2}\,.
\end{equation}

This gives the holomorphic part of the vacuum block $\mathcal{F}^{\Gamma(z_{c_1},z_{c_2})}_0$ as 
\begin{align}
    &\frac{\left(z_1 z_2\right)^{-h}}{(4 \rho^2)^{-2h}} \left[\left( \left(\frac{z_{c_1}}{z_{c_2}}\right)^{1/2} - \left(\frac{z_{c_1}}{z_{c_2}}\right)^{-1/2} \right) \left( \left(\frac{z_1}{z_{c_1}}\right)^{i \rho / 2} - \left(\frac{z_1}{z_{c_1}}\right)^{-i \rho / 2} \right) \left( \left(\frac{z_2}{z_{c_2}}\right)^{i \rho / 2} - \left(\frac{z_2}{z_{c_2}}\right)^{-i \rho / 2} \right)\right.\nonumber\\
    +& 2 i \rho \left( \left(\frac{z_{c_1}}{z_{c_2}}\right)^{1/2} + \left(\frac{z_{c_1}}{z_{c_2}}\right)^{-1/2} \right) \left( \left(\frac{z_1 z_{c_2}}{z_{c_1} z_2}\right)^{-i \rho / 2} - \left(\frac{z_1 z_{c_2}}{z_{c_1} z_2}\right)^{i \rho / 2} \right)\nonumber\\
    +&\left. \rho^2 \left( \left(\frac{z_{c_1}}{z_{c_2}}\right)^{1/2} - \left(\frac{z_{c_1}}{z_{c_2}}\right)^{-1/2} \right) \left( \left(\frac{z_1}{z_{c_1}}\right)^{i \rho / 2} + \left(\frac{z_1}{z_{c_1}}\right)^{-i \rho / 2} \right) \left( \left(\frac{z_2}{z_{c_2}}\right)^{i \rho / 2} + \left(\frac{z_2}{z_{c_2}}\right)^{-i \rho / 2} \right)\right]^{-2h}
\end{align}
The $(z_1 z_2)^{-h}$ factor is the Jacobian factor to be canceled out by going from cylindrical to planar coordinates. We use the coordinates $z_1=e^{\tau_1+i \theta/2}$, $z_2=e^{\tau_2-i \theta/2}$, $z_{c_i}=e^{i w_i}$. In fact, because we are mostly interested in antipodal points $\theta=\pi$, it is useful to parameterize the crossing points using $\sigma=w_1+w_2$ and $\delta=w_2-w_1+\pi$. The anti-holomorphic part is given by using conjugate coordinates $\bar{z}$. We can then analytically continue to real time $\tau_i=it_i$. We have the vacuum block in the channel labeled by $\sigma,\delta$ as
\begin{equation}
    \abs{\mathcal{F}^{\Gamma(z_{c_1},z_{c_2})}_0}^2=\frac{\rho^{4\Delta}}{\qty(\mathcal{B}_+(t_1,t_2,\sigma,\delta)\mathcal{B}_-(t_1,t_2,\sigma,\delta))^{\Delta}}
\end{equation}
where
\begin{align}
    \mathcal{B}_\pm(t_1,t_2,\sigma,\delta)=&\cos\frac{\delta}{2} \left[ (-1 + \rho^2) \cosh\frac{\rho(t_2 - t_1 \pm \delta)}{2} + (1 + \rho^2) \cosh\frac{\rho(t_2 + t_1\pm \sigma)}{2} \right]\\
    &\pm 2 \rho \sin\frac{\delta}{2} \sinh\frac{\rho(t_2 - t_1 \pm \delta)}{2}\,.
\end{align}
Taking the sum in \eqref{eq:sum-channels} to be the integral over $w_1,w_2$ with measure $dw_1dw_2$, the full correlator is therefore the integral 
\begin{equation}\label{eq:int-over-channels}
    G\qty(t_1,\frac{\pi}{2};t_2,-\frac{\pi}{2})=\frac{1}{2}\int_{-\pi}^\pi d\delta\int_{-\infty}^{\infty} d\sigma\frac{\rho^{4\Delta}}{\qty(\mathcal{B}_+(t_1,t_2,\sigma,\delta)\mathcal{B}_-(t_1,t_2,\sigma,\delta))^{\Delta}}
\end{equation}

The bounds of the integral require some explanation. We stated that the correlator is given by a sum over all possible monodromies. However, as shown in Fig. \ref{fig:crossing-pts}, there are in fact contraction paths that are topologically distinct despite having the same crossing points. The integration range in \eqref{eq:int-over-channels} represents the contraction paths that are deformations of the one in the figure on the left, without crossing either of the two points at which the probe fields are inserted. More precisely, it is required that $w_2+2\pi>w_1>w_2$, which leads to $-\pi<\delta<\pi$, whereas there are no restrictions on $\sigma$. The monodromy method appears to also account for the paths of other topologies, but with $\delta$ taking values outside the range described above. In the prescription \eqref{eq:sum-channels}, one should consider all channels, including these non-trivial paths, but in this work we will only focus on the sector given by $\abs{\delta}<\pi$. This is because, as we will see below, there are some subtleties if one is to consider the region $\abs{\delta}>\pi$ seriously, and it turns out that the ``trivial'' sector alone is enough for us obtain the results on the bulk cone singularities.

\section{Light cone and bulk cone singularities}
\label{sec:analysis}
In this section, we analyze the integral \eqref{eq:int-over-channels} and show that it reproduces the bulk cone singularities expected from Section \ref{sec:bulk}. Along the way, we will also analyze the more familiar light cone singularity. We will study these singularities from two perspectives. Much like the analysis in \cite{Anous:2017tza}, one can consider the integral \eqref{eq:int-over-channels} in the saddle point approximation at large $\Delta$ (while still keeping $\Delta\ll c$). However, the bulk cone singularity is a general feature independent of $\Delta$ and one should not need to resort to the saddle point approximation to see them. Indeed, one can also see these singularities just by studying the properties of \eqref{eq:int-over-channels}, which we will do in Section \ref{sec:direct-integral}. In Section \ref{sec:saddle}, we will discuss the saddle points.

\subsection{Direct analysis of the integral over channels}
\label{sec:direct-integral}
We would like to understand how the integral \eqref{eq:int-over-channels} gives rise to the light cone and bulk cone singularities. The key to this is to first understand the possible singularities of the integrand, which occur when one of $\mathcal{B}_\pm(t_1,t_2,\sigma,\delta)$ vanishes.\footnote{Although the integral involves a non-compact direction in $\sigma$, the integrand decays exponentially for large $\abs{\sigma}$ at any fixed $\abs{\delta}<\pi$. On the boundary $\abs{\delta}=\pi$, the integrand is constant. So the only way the integral diverges is through singularities of the integrand.} Fixing $t_1,t_2$, we denote the loci on the $(\delta,\sigma)$-plane at which $\mathcal{B}_\pm(t_1,t_2,\sigma,\delta)$ vanishes by $L_\pm(t_1,t_2)$. Note that the definition of $L_\pm(t_1,t_2)$ depends on the range of $\delta,\sigma$ that we consider. As explained in the previous section, the integration range is $\abs{\delta}\leq\pi$, so we will mostly consider $L_\pm(t_1,t_2)$ in this range and in a neighborhood slightly beyond it.

\begin{figure}[H] 
 \begin{center}                      
      \includegraphics[width=5in]{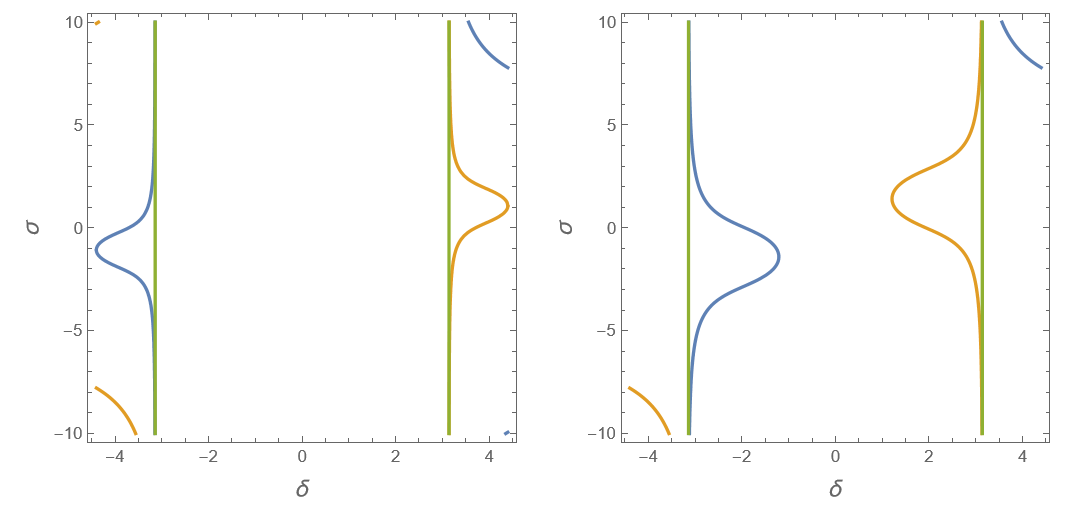}
      \caption{The loci $L_\pm$ in the $(\delta,\sigma)$-plane at different $t_2,t_1$. The region between the green lines represent the boundary of the integration region. The blue and yellow lines represents $L_+$ and $L_-$. On the left, we have $t_2-t_1<\pi$ and the integrand has no singularities in the integration region; while $t_2-t_1>\pi$ on the right, and the integrand is singular at the loci and need an $i\epsilon$ prescription. Note that there are disconnected components for both $L_\pm$, which is related to the fact that $\mathcal{B}_\pm=0$ has infinitely many solutions if we consider $\delta$ beyond the range of integration.}
  \label{fig:singularity-loci}
  \end{center}
\end{figure}

Due to the symmetry that relates the two factors $\mathcal{B}_\pm$, it suffices to look at the singularity locus coming from one of them. Setting $\mathcal{B}_+=0$, we can solve for $L_+$ as
\begin{equation}\label{eq:locus}
    \sigma=-(t_1+t_2)\pm \frac{2}{\rho}\arccosh\qty[\frac{(1-\rho^2)\cosh\frac{\rho(t_2-t_1+\delta)}{2}-2\rho\tan\frac{\delta}{2}\sinh\frac{\rho(t_2-t_1+\delta)}{2}}{1+\rho^2}]\,.
\end{equation}
The two branches are joined together smoothly at $\sigma=-(t_1+t_2)$. $L_-$ is given by taking $\sigma\to -\sigma$ and $\delta\to -\delta$. Real values of $\sigma$ are only obtained when the argument inside the $\arccosh$ is greater than 1. Within the integration region $\abs{\delta}<\pi$, this never happens if and only if $t_2-t_1<\pi$. In fact, at exactly the light cone time $t_2-t_1=\pi$, $L_\pm$ are given by $\delta=\mp \pi$, which are the boundary of the integration region. $L_\pm$ enters the region if $t_2-t_1>\pi$. 

When the line of singularity is within the integration region, the integral is defined by an $i\epsilon$ prescription given by the ordering of the operators, i.e. to get the ordering in \eqref{eq:gen-corr}, we need the Euclidean ordering $\tau_1<0<\tau_2$.\footnote{$\tau_1,\tau_2$ needs to be of opposite signs for the entire locus of singularity to be shifted in one direction in the complex plane, i.e. if we take $t_1\to t_1+i\epsilon$, $t_2\to t_2+i\gamma\epsilon$ and expand $\mathcal{B_+}$ to leading order in $\epsilon$, we must have $\gamma<0$ in order for the coefficient of $\epsilon$ to have a consistent sign throughout $L_+$.}
So that \eqref{eq:int-over-channels} is in fact
\begin{equation}\label{eq:int-epsilon}
    \lim_{\epsilon\to 0}\frac{1}{2}\int_{-\pi}^\pi d\delta\int_{-\infty}^{\infty} d\sigma\frac{\rho^{4\Delta}}{\qty(\mathcal{B}_+(t_1,t_2,\sigma,\delta)+i \epsilon)^\Delta(\mathcal{B}_-(t_1,t_2,\sigma,\delta)-i \epsilon)^{\Delta}}
\end{equation}
with $\epsilon>0$. At generic times, this shifts the singularities away from the integration contour and gives a well-defined integral.\footnote{This analysis focuses on the integration region $\abs{\delta}<\pi$, but as mentioned in the last section, there are in fact monodromies with distinct topology that is parameterized by $\delta$ beyond this range. The argument above clearly shows that even before the light cone time $t_2-t_1<\pi$, $L_\pm$ is supported outside of $\abs{\delta}<\pi$. This would imply that the integral would need an $i\epsilon$ prescription even before $t_2-t_1<\pi$. This is certainly unexpected and we hope to understand this further in future work.} However, as mentioned, the singularity of the integrand at the light cone time $t_2-t_1=\pi$ coincide exactly with the boundary of the region of integration. In this case, the integral is infinite regardless of the $i\epsilon$ prescription, thus giving rise to the light cone singularity.\footnote{More precisely, at the light cone time $t_2-t_1=\pi$, the $\delta$-integral near say $-\pi$ is proportional to
\begin{equation}
    \int_{-\pi}^{\delta_0} \frac{d\delta}{(\delta+\pi+i\epsilon)^\Delta}\sim \frac{(i\epsilon)^{1-\Delta}}{\Delta-1}\,,
\end{equation}
which diverges as $\epsilon\to 0$.} In fact, it is satisfying to see that the light cone singularity is associated with the boundary $\delta=\pm\pi$ because this is exactly when the crossing points degenerate $z_{c_1}=z_{c_2}$. As mentioned, this contribution is equivalent to the monodromy for the contour that does not cross the shell and is therefore no different from a heavy-heavy-light-light four-point correlator, which of course has the same light cone singularity.

\begin{figure}[H] 
 \begin{center}                      
      \includegraphics[width=5in]{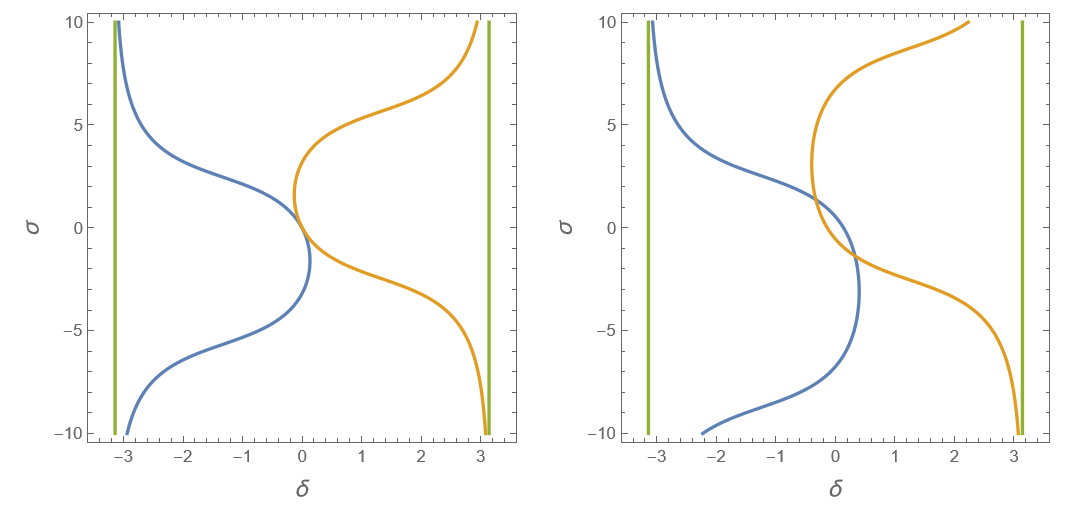}
      \caption{Left: At exactly the bulk cone time $t_2=t_C(t_1)$, $L_\pm$ intersect at the origin tangent to each other. Right: Past the light cone time $t_2>t_C(t_1)$, $L_\pm$ has two generic intersections. In cases where no singularities are expected from bulk calculations, i.e. when $\rho>1$ or when $\rho<1$ but $t_1$ is too late for a bulk cone singularity to exist, $L_\pm$ never intersect.}
  \label{fig:pinch}
  \end{center}
\end{figure}

We now turn to the bulk cone singularity. Once the time difference is greater than the light cone time $t_2-t_1>\pi$, $L_\pm$ bulge inward. Depending on the values of $\rho$, $t_1$ and $t_2$, the originally non-intersecting $L_\pm(t_1,t_2)$ may or may not intersect eventually as $t_2-t_1$ increases. In the case where the loci never intersect, it is clear that no bulk cone singularity occurs since the $i\epsilon$ prescription gives a finite integral. The case that they eventually intersect is more interesting. Suppose we fix $t_1$, there is a critical time $t_2=t_c(t_1)$ at which the two loci meet for the first time tangentially at one point, and after which they have a transverse intersection at two points (see Fig. \ref{fig:pinch}). It can be checked that the location at which $L_\pm(t_1,t_c(t_1))$ meets tangentially is always at $\delta=0=\sigma$.\footnote{This is true since we set the two points to be antipodal $\theta=\pi$, but it would not be the case if we are considering generic angular separations. However, the same bulk cone singularity occurs in those case and they in fact arise from the same mechanism described here.} Given this fact, we can just plug in $\delta=0,~\sigma=0$ in \eqref{eq:locus} to solve for $t_2=t_c(t_1)$
\begin{equation}\label{eq:BC-time-CFT}
    t_c(t_1)=\frac{1}{\rho} \log\qty(\frac{\tanh\frac{\rho t_1}{2}-\rho^2}{\tanh\frac{\rho t_1}{2}+\rho^2})\,,
\end{equation}
which is exactly the bulk cone time \eqref{eq:BC-time}!\footnote{On the CFT, we have time in units of $l$.} This explains the absence of bulk cone singularities in the case of $\rho>1$, where \eqref{eq:BC-time-CFT} is complex and $L_\pm$ never intersects. It then remains to be seen that in cases with $\rho<1$, the integral diverges when $L_\pm$ meets tangentially, which would show that the CFT correlator exhibits the bulk cone singularity expected from the bulk.

To see this, we zoom in to the origin since the divergent behavior is local to $\sigma=\delta=0$ at the bulk cone singularity. By a rotation, we can switch to orthogonal coordinates $\sigma',\delta'$ such that $\delta'=0$ coincides with the tangential line at which $L_\pm$ meets at the origin. Near the bulk cone time $t_2=t_c(t_1)$, in these coordinates, the integral takes the form
\begin{equation}\label{eq:origin-near-BC}
    \sim\int d\sigma'\int d\delta' \frac{1}{(\delta' + A_1(\sigma'-A_2)^2 + A_3-i\epsilon)^{\Delta} (\delta' - A_1 (\sigma'+A_2)^2 - A_3+i\epsilon)^\Delta}\,,
\end{equation}
where $\sim$ means that we have expanded around the origin and dropped factors unrelated to the divergence. As for the parameters, $A_1$ is non-vanishing and negative near $t_2=t_c(t_1)$, while $A_2$ and $A_3$ both vanish when $t_2=t_c(t_1)$. Moreover, $A_3<0$ when $t_2<t_c(t_1)$ and $A_3>0$ when $t_2>t_c(t_1)$. See Appendix \ref{app:expansion} for more details. This models the quadratic behavior of $L_\pm$ near the point where they meet tangentially. Integrating $\delta'$ gives
\begin{equation}\label{eq:sigma-integral}
    \sim\int d\sigma'\frac{1}{(A_1 (\sigma'^2+A_2^2) + A_3-i\epsilon)^{2\Delta-1}}\,.
\end{equation}
Before the bulk cone singularity $t_2<t_c(t_1)$, $A_1$ and $A_3$ have the same signs and the integral can be done without the $i\epsilon$. However, for $t_2>t_c(t_1)$, there are zeroes in the denominator, which represent the crossings of the two loci as shown in the right figure in Fig. \ref{fig:pinch}. Evaluating \eqref{eq:sigma-integral} as
\begin{equation}
    \int d\sigma'\frac{1}{\qty[A_1\qty(\sigma'-\sqrt{\frac{A_3+A_1A_2^2}{-A_1}}+i \epsilon)\qty(\sigma'+\sqrt{\frac{A_3+A_1A_2^2}{-A_1}}-i \epsilon)]^{2\Delta-1}}\,,
\end{equation}
we find that this is finite given that $A_2\neq 0$ and $A_3\neq 0$ because the singularities at $\delta' + A_1(\sigma'-A_2)^2 + A_3=0$ are taken into account by the $i\epsilon$ prescription. However, due to the relative sign between the two $i\epsilon$'s, there is a pinch singularity in the limit $A_2,A_3\to 0$ which happens when $t_2=t_c(t_1)$. Thus, we have shown the divergence of the large $c$ CFT correlator at the bulk cone time.

\subsection{Saddle point analysis}
\label{sec:saddle}
We will now consider \eqref{eq:int-over-channels} in a saddle point analysis for $\Delta \gg 1$. We are only interested in the leading order contribution, i.e. $G\approx e^S$ where $S=O(\Delta)$. The critical points are given by
\begin{equation}\label{eq:sigma-eqn}
    0=\partial_\sigma (\mathcal{B}_+\mathcal{B}_-)=\frac{\rho (1 + \rho^2)}{2} \cos\frac{\delta}{2} 
    \left(
    -\mathcal{B}_+ \sinh\frac{\rho (t_1 + t_2 - \sigma)}{2} 
    + \mathcal{B}_- \sinh\frac{\rho (t_1 + t_2 + \sigma)}{2} 
    \right)\,,
\end{equation}
\begin{align}\label{eq:delta-eqn}
    0=&\partial_\delta (\mathcal{B}_+\mathcal{B}_-)\nonumber\\
    =&
    \frac{1}{2} (1 + \rho^2) \Bigg[
    \mathcal{B}_+ \Bigg(
        \left(\cosh\frac{\rho(t_2 - t_1 - \delta)}{2} 
        - \cosh\frac{\rho (t_1 + t_2 - \sigma)}{2}\right) 
        \sin\frac{\delta}{2}
        - \rho 
        \sinh\frac{\rho(t_2 - t_1 - \delta)}{2}
        \cos\frac{\delta}{2}
    \Bigg)\nonumber
    \\
    &+\mathcal{B}_- \Bigg(
        \left(\cosh\frac{\rho(t_2 - t_1 + \delta)}{2} 
        - \cosh\frac{ \rho (t_1 + t_2 + \sigma)}{2}\right) 
        \sin\frac{\delta}{2}
        + \rho
        \sinh\frac{\rho(t_2 - t_1 + \delta)}{2}
        \cos\frac{\delta}{2}
    \Bigg)
    \Bigg]\,.
\end{align}
While we have so far described various quantities at fixed $t_1$ and varied $t_2$, the structure of the equations indicate that it is often more natural to fix $\bar{t}=t_1+t_2$ and vary the time difference $\delta t=t_2-t_1$. In these coordinates, the constraints $t_1<0$ and $t_2>0$ becomes $\delta t>\bar{t}$ and $\delta t>-\bar{t}$.

\subsubsection*{Before light cone time $\delta t<\pi$}
We will first consider the saddle point analysis for times before the light cone time $\delta t<\pi$, where the integrand is purely real. In this case, the saddle point method simply means finding the maxima within the integration region.

Note that the first equation is always satisfied for $\delta=\pm \pi$, which is the boundary of our integration region and correspond to the two connection points degenerating to one point, i.e. the contribution without crossing the shell. This implies the integrand on the entire boundary takes the constant value
\begin{equation}\label{eq:bdy-saddle}
    \qty(\frac{\rho^2}{4 \sinh\left(\frac{1}{2} (\pi - \delta t) \rho\right) \sinh\left(\frac{1}{2} (\pi+\delta t) \rho\right)})^{\Delta}\,.
\end{equation}
Before the light cone time, the equation for $\delta$ has two solutions on both $\delta=\pm \pi$. As $\delta t$ increases towards $\pi$, the two solutions on $\delta=-\pi$ approaches $\sigma=-\bar{t}$ and those on $\delta=\pi$ approaches $\sigma=\bar{t}$. Indeed, one can check that both of these limiting points solves the saddle equations at exactly $\delta t=\pi$.

\begin{figure}[H] 
 \begin{center}                      
      \includegraphics[width=5in]{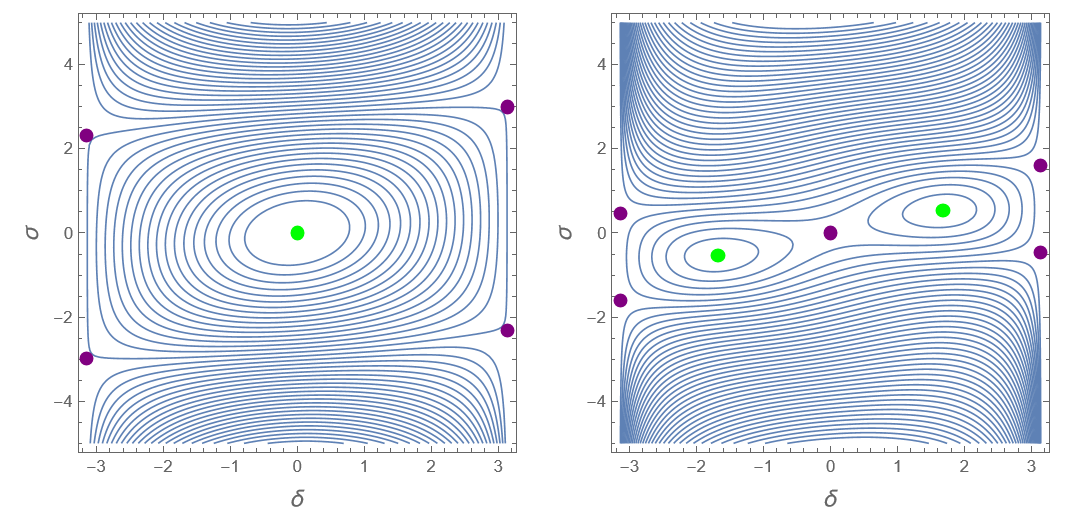}
      \caption{The critical points of the integrand on the $(\delta,\sigma)$-plane. Dominant saddles are labeled green. Left: for small $\delta t$, or precisely, when both \eqref{eq:split-time} and \eqref{eq:small-t-bar} are satisfied, we have five critical points, including four on the boundary and one at the origin. Right: Beyond the range described above, we have seven critical points.}
  \label{fig:real-saddles}
  \end{center}
\end{figure}

Now turning to other saddles, one can check that there is always a critical point at the origin $\delta=0=\sigma$ with value
\begin{equation}\label{eq:origin-saddle}
    \frac{\rho^{4\Delta}}{\qty((1+\rho^2)\cosh\frac{\rho\bar{t}}{2}+(-1+\rho^2)\cosh\frac{\rho \delta t}{2})^{2\Delta}}\,.
\end{equation}
At small enough time separations $\delta t$, these five solutions are all the critical points in the integration region, but as $\delta t$ increases, there are two additional critical points that split off from the one at the origin and eventually move to the limiting points $\delta=-\pi,\sigma=-\bar{t}$ and $\delta=\pi,\sigma=\bar{t}$. Since each critical point is given by the intersection between the loci satisfying both \eqref{eq:delta-eqn} and \eqref{eq:sigma-eqn}, we can understand the splitting of critical points at the origin as the two loci becoming tangent to each other at their intersection, after which there are three intersections, so we can find the time of the splitting by matching the slopes of each of the loci at the origin. This gives the time difference for the splitting at fixed $\bar{t}$\footnote{In fact, the slopes of the loci coincide again at the bulk cone time.}
\begin{equation}\label{eq:split-time}
    \delta t_{split}=\frac{2}{\rho} \log\left(\frac{1}{2} \left(2 + 2 \rho^2 + \sqrt{2 + 8 \rho^2 + 4 \rho^4 - 2 \cosh\rho \bar{t} }\right) \sech\frac{\rho\bar{t} }{2}\right)\,.
\end{equation}
Recall that it is necessary to impose $\delta t_{split}>\pm\bar{t}$, which is not always satisfied by \eqref{eq:split-time}. Solving for this condition gives
\begin{equation}\label{eq:small-t-bar}
    \frac{\log\left(1 + 2 \rho^2 - 2 \sqrt{\rho^2 (1 + \rho^2)}\right)}{\rho}<\bar{t}<\frac{\log\left(1 + 2 \rho^2 + 2 \sqrt{\rho^2 (1 + \rho^2)}\right)}{\rho}\,,
\end{equation}
i.e. this qualifies what was meant by ``small enough'' $\delta t$, since it is only for this range of $\bar{t}$ where there exist small enough $\delta t>\pm \bar{t}$ satisfying $\delta t_{split}>\delta t>\pm \bar{t}$ such that one starts with only five critical points. If one fixes $\bar{t}$ to be out of this range, there are always seven critical points regardless of $\delta t$.

We can now discuss the dominant saddles. In cases where $\bar{t}$ satisfy \eqref{eq:small-t-bar} and $\delta t<\delta t_{split}$, one only needs to consider compare the saddle at $\delta=0=\sigma$ \eqref{eq:origin-saddle} with the boundary value \eqref{eq:bdy-saddle} and one finds that the saddle at the origin is dominant. As $\delta t$ increases beyond this range, the boundary saddle values eventually becomes larger than that at the origin but it is always the case that the two new saddles dominate over the boundary saddles and the origin saddle. Similarly, when $\bar{t}$ is beyond the range of \eqref{eq:small-t-bar}, the two saddles that are moving away from the origin always dominates.

Next, we will consider the light cone singularity in the saddle point approximation. As mentioned, when we approach the light cone time $\delta t=\pi$, the two saddles that always dominate in this regime approach the points $\delta=-\pi,\sigma=-\bar{t}$ and $\delta=\pi,\sigma=\bar{t}$. It is difficult to analytically describe the trajectory of these saddles points since the saddle equations are quite complicated, so we will consider near light cone behavior in the special case of $\bar{t}=0$ and we expect similar results in the general case. $\bar{t}=0$ greatly simplifies the saddle equations since \eqref{eq:delta-eqn} is solved by $\sigma=0$. We expand the remaining equation near the endpoint, say $\delta=-\pi,\sigma=-\bar{t}$ to $O((\delta+\pi)^2)$ and solve for $\delta t$, which we can then invert to get
\begin{equation}
    \delta=\pi-2\qty(\frac{1}{\rho}\tanh\frac{\rho \pi}{2})^{1/2}\sqrt{\pi-\delta t}
\end{equation}
Putting this into the integrand then gives leading light cone singularity as
\begin{equation}
    2\qty(\frac{\rho}{2\sinh(\rho\pi)(\pi-\delta)})^\Delta\,.
\end{equation}

\subsubsection*{Past the light cone time $\delta t>\pi$}
The dominant saddles before the light cone time approaches $(\delta,\sigma)=(\pm\pi,\pm\bar{t})$ where it merges with the other critical points on the boundary and diverges. After this, the saddles move into the complex plane. Unlike when the points are only spacelike, the integrand is no longer real, since the integral is only convergent by virtue of the $i\epsilon$ prescription. Therefore, we use the method of steepest descent to determine whether a complex saddles contribute the the integral.

\begin{figure}[H] 
 \begin{center}                      
      \includegraphics[width=2.3in]{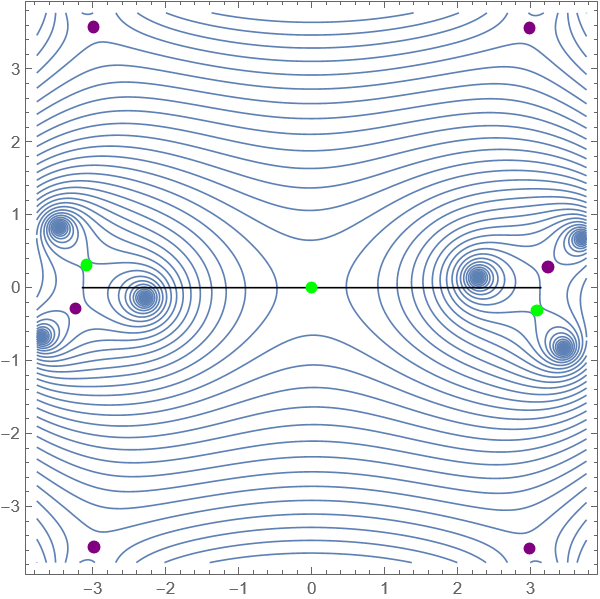}
      \caption{Saddles in the complex $\delta$-plane at $\sigma=0$ when $\bar{t}=0$. The straight line is the original integration contour. The points are saddles and the contributing saddles are labeled gree. Here the singularities on the contour are shifted away by turning on $i\epsilon$.
      There are saddles further away from the contour $\textnormal{Im}(\delta)=0$, but they do not contribute. There are pairs of saddles near the boundary of the integration region and only one within the pair contributes. The saddle at the origin also contributes and is responsible for the bulk cone singularity.}
  \label{fig:complex-saddles}
  \end{center}
\end{figure}

For simplicity, we will again restrict to the simpler case of $\bar{t}=0$, where the complex saddles lie on the complex $\delta$-plane at $\sigma=0$.\footnote{The more general case has similar features but the complex saddles lie on a surface in $\mathbb{C}^2$ that reaches $(\delta,\sigma)=(\pm\pi,\pm\bar{t})$ instead.} The saddle points past the light cone time are shown in Fig. \ref{fig:complex-saddles}. Note that the integrand shown has a small $\epsilon$ turned on. We still have the saddle at the origin, and the other saddles become conjugate pairs in the limit $\epsilon\to 0$. The saddles further away from $\textnormal{Im}(\delta)=0$ do not contribute since the original contour cannot be deformed to the steepest descent contours for these saddles. The pairs of saddles near the $\delta=\pm \pi$ are symmetric in the vanishing $\epsilon$ limit, but the $i\epsilon$ prescription in fact picks out one saddle out of the pair. Similarly, the saddle at the origin appears to have a steepest descent contour orthogonal to the original contour, but the $i\epsilon$ prescription allows the saddle to contribute. Therefore, we have three saddles to consider.

\begin{figure}[H] 
 \begin{center}                      
      \includegraphics[width=3in]{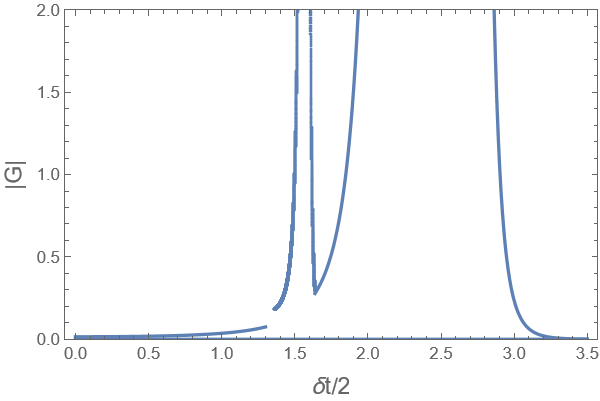}
      \caption{This figure shows the modulus of the correlator evaluated on the saddles, as a function of $\delta t/2$ at $\bar{t}=0$. The parameters are $\Delta=3$, $\rho=0.7$. Note that there are two times where the correlator is divergent, with the earlier time as the light cone time $\delta t=\pi$ and the later time as the bulk cone time. The discontinuity is at $\delta t_{split}$ where the saddle becomes degenerate and the saddle point method is unreliable.}
  \label{fig:corr-on-saddles}
  \end{center}
\end{figure}

For $\delta t$ close to $\pi$, the pairs of saddles dominates since they are in the vicinity of the singularity of the integrand. As $\delta t$ increases further, the saddle at the origin becomes dominant and is related to the bulk cone singularity as it is pinched by $L_\pm$ in the case of $\rho<1$, otherwise its contribution remains finite. Fig. \ref{fig:corr-on-saddles} shows the correlator evaluated on the saddles discussed. Finally, we state the leading singularity from the saddle at the origin \eqref{eq:origin-saddle} as $t_2$ approaches the bulk cone time $t_c$ at fixed $t_1$
\begin{equation}
    \qty[\frac{\rho^2(1+\rho^2)}{2}\qty((1-\rho^2)\cosh\rho t_1-1)]^{\Delta} \frac{1}{(t_c-t_2)^{2\Delta}}\,.
\end{equation}

\section{Discussion}
\label{sec:disc}
We have studied a family of time-reversal symmetric spacetimes given by a heavy operator acting on the vacuum and found a transition in the causal structure at $r_+=l$. Below the threshold, the bulk spacetime contains radial null geodesics sent from the boundary that can return to the boundary at an antipodal point in finite time, whereas radial null geodesics always fall inside the black hole above the transition. These can be detected on the boundary as bulk cone singularites in the two-point function in the state dual to such spacetimes.

We then calculated a CFT$_2$ correlator corresponding to the two-point function in that geometry using large $c$ conformal blocks. These are calculated using the monodromy method and we used a sum over channels to obtain the Lorentzian result. The correlator is then analyzed directly as an integral and through the saddle point approximation for large $\Delta$. In both cases, we find that a singularity in the CFT correlator at exactly the bulk cone time expected by bulk calculations.

\subsubsection*{A von Neumann algebra interpretation of the transition}
In this work, we study the transition in the causal structure via bulk cone singularities but we note that there is a different way to characterize the effect using von Neumann algebras. The authors of \cite{Leutheusser:2022bgi} proposed a duality between boundary subalgebras and bulk subregion, extending the usual framework of entanglement reconstruction. In the more familiar entanglement reconstruction, one associates the large $N$ algebra of observables on the boundary with the algebra of observables in the entanglement wedge. \cite{Leutheusser:2022bgi} studied the smaller single trace algebra, which contains only generalized free fields, and proposed that the single trace algebra on the entire boundary is dual to the causal completion of the algebra of the union of all causal wedges.

This is relevant to the geometries that we study in this work since below the threshold $r_+<l$, the union of all causal wedges contains an entire Cauchy slice, which allows for causal evolution. The causal completion of the algebra is therefore the algebra of observables in the entire spacetime in the sense of QFT in curved spacetime. However, above the threshold, the union of causal wedges does not include an entire Cauchy slice and the region in the overlap of the black hole and white hole is missing. Therefore, even under causal completion, one only obtains a subalgebra for the entire geometry. Using standard results of QFT in curved spacetimes, we know that the algebra in the entire spacetime is type I and that of a causally complete subregion that is not the entire space time is type III. This then directly translates to the statement that the generalized free field algebra $\mathcal{A}$ of the form
\begin{equation}\label{eq:corr}
    \langle 0\vert V^{\dagger}(t=0)OV(t=0)\vert 0\rangle\,,
\end{equation}
with $O\in \mathcal{A}$ has a transition in its type when the dimension of $V$ crosses the threshold.

This raises the question of whether there is the connection between bulk cone singularities and the type of single trace algebra. Since the algebra is a generalized free field, the two-point function that we calculated is sufficient to deduce the type of algebra via the Araki-Woods construction (see e.g. \cite{Faulkner:2022ada,Gesteau:2024dhj} for recent reviews). In particular, divergences such as the bulk cone singularity can play an important role in the spectral analysis of the two-point function that determines the type of the algebra. It would be an interesting challenge to understand this possible connection further.

\subsubsection*{Bulk cone singularities from non-radial geodesics}
As mentioned in the introduction, the more well-studied bulk cone singularities in black hole backgrounds arises from null geodesics that carry angular momentum and wind around the black hole. One would expect these to appear in the family of spacetimes that we consider in this work. However, the calculation presented here, which successfully captures the behavior of the radial bulk cone singularity, does not seem to detect those bulk cone singularities at all. In fact, even the usual light cone singularity should have a periodic structure simply because of the compact nature of the spatial slice, and yet we do not see this structure in the CFT calculation. A possible answer to this puzzle is that we did not in fact sum over all possible monodromies. As was pointed out at the end of Section \ref{sec:cft} and in Fig. \ref{fig:crossing-pts}, there are contraction paths that are topologically distinct from the ones that we considered and should also be accounted for. It would not be surprising that these ``winding'' paths turn out to be related to singularities coming from geodesics that wind around the black hole. The most straightforward way to incorporate these paths is to simply extend the integral over channels beyond the ``non-winding'' paths. However, as mentioned in Section \ref{sec:direct-integral}, due to the singularities in the integrand beyond the range that we considered, there might be subtleties in defining the integral. It would be illuminating to understand this further.

\subsubsection*{Further CFT understanding of bulk cone singularities}
Although we have given a CFT calculation that successfully reproduces bulk cone singularities, it uses large $c$ CFT technology that does not provide easy physical interpretation of the singularities. In particular, in the sum over channels, the singularities $L_\pm$ in various channels at Lorentzian times are crucial in producing the bulk cone singularities but they do not seem to have an obvious CFT interpretation. In fact, we understand the bulk cone singularity through the collision of $L_\pm$. This geometric point of view seems to just be an artifact of the continuum limit. When there are only finitely many operator insertions, the analogue of $L_\pm$ would be a discrete set of paths and the interpretation of possible bulk cone singularities would be very different. Therefore, it would be desirable to understand the calculation further in cases where it is not necessary to take a continuum limit. A related direction is to understand how the bulk cone singularity emerges from a large $c$ or large $N$ limit. It is interesting to explore possible finite $N$ models where explicit calculations can be done and observe a bulk cone singularity appear in the limit. Finally, our original bulk calculation is independent of dimension, but our CFT calculations is only applicable for AdS$_3$/CFT$_2$. In the case of higher dimensions, a natural guess for the calculation that would give rise to the bulk cone singularity is to consider the global conformal blocks of the stress tensor.

\section*{Acknowledgements}
We would like to thank David Grabovsky, Jesse Held, Sam Leutheusser, Robinson Mancilla, Don Marolf, Leonel Queimada, Jiuci Xu and Ying Zhao for helpful discussions; and especially Gary Horowitz for his guidance and patience throughout the completion of this project. This research was supported in part by NSF Grant PHY-2408110.

\appendix
\section{Expansion of \eqref{eq:int-epsilon} near bulk cone singularity}
\label{app:expansion}
In this appendix, we obtain \eqref{eq:origin-near-BC} from \eqref{eq:int-epsilon} by expanding near the origin in the $(\delta,\sigma)$-plane. As in Section \ref{sec:saddle}, we use variables $\delta t=t_2-t_1$ and $\bar{t}=t_1+t_2$. We describe the approach to the bulk cone singularity at fixed $\bar{t}$ as $\delta t\to\delta t_c$. We consider the case $\rho<1$ throughout this appendix, so $\delta t_c$ is always real. As mentioned before \eqref{eq:origin-near-BC}, we rotate to new coordinates $(\delta',\sigma')$ in order to align $\delta'=0$ with the tangent line at which $L_\pm$ first intersect. This is achieved by $(\delta',\sigma')=(\delta \cos\phi+\sigma\sin\phi,\sigma \cos\phi-\delta\sin\phi)$, where $\tan\phi=\frac{\sinh(\rho\bar{t}/2)}{\sinh(\rho\,\delta t/2)}$.  Expanding $\mathcal{B}_+(t_1,t_2,\sigma,\delta)$ to second order in $\delta',\sigma'$ gives
\begin{equation}\label{eq:second-ord}
    \mathcal{B}_+(t_1,t_2,\sigma,\delta)\approx c_0+c_1\delta'+c_2\sigma'^2+c_3\delta'^2+c_4 \delta'\sigma'\,,
\end{equation}
where
\begin{align}
    c_0&=(1+\rho^2)\cosh\frac{\rho \bar{t}}{2}+(-1+\rho^2)\cosh\frac{\rho \delta t}{2}\\
    c_1&=\frac{\rho(1+\rho^2)}{2}\sqrt{\frac{D\qty(\bar{t},\delta t)}{2}}\\
    c_2&=\frac{1+\rho^2}{4D\qty(\bar{t},\delta t)}
    \qty[(\rho^2+1)\cosh\frac{\rho\,\delta t}{2}\sinh^2\frac{\rho\bar{t}}{2}
    +\cosh\frac{\rho\bar{t}}{2}\qty(\rho^2\sinh^2\frac{\rho\delta t}{2}
    -\sinh^2\frac{\rho\bar{t}}{2})]\\
    c_3&=\frac{1+\rho^2}{4D\qty(\bar{t},\delta t)}
    \qty[(\rho^2+1)\cosh\frac{\rho\,\delta t}{2}\sinh^2\frac{\rho\delta t}{2}
    +\cosh\frac{\rho\bar{t}}{2}\qty(\rho^2\sinh^2\frac{\rho\bar{t}}{2}
    -\sinh^2\frac{\rho\delta t}{2})]\\
    c_4&=\frac{(1+\rho^2)^2}{2D\qty(\bar{t},\delta t)}\sinh\frac{\rho\bar{t}}{2}\sinh\frac{\rho\delta t}{2}\qty(\cosh\frac{\rho\bar{t}}{2}-\cosh\frac{\rho\delta t}{2})\,,
\end{align}
with $D\qty(\bar{t},\delta t)=-2+\cosh\rho\bar{t}+\cosh\rho\,\delta t$. The important point here is that exactly at the bulk cone time \eqref{eq:BC-time-CFT}, $c_0$ vanishes, whereas the other coefficients are non-vanishing.

The singularity locus $L_+$ is given by $\mathcal{B}_+=0$. We may set \eqref{eq:second-ord} to 0 and solve for $\delta'$ as a function of $\sigma'$ up to second order. Due to the fact that we did an expansion to get \eqref{eq:second-ord}, this yields two solutions but only one represents $L_+$. We therefore obtain
\begin{equation}
    \mathcal{B}_+(t_1,t_2,\sigma,\delta)\approx c_1(\delta'-(A_1(\sigma+A_2)^2+A_3))\,,
\end{equation}
where $A_i$, $i=1,\ldots,3$ can be expressed in terms of $c_j$, $j=0,\ldots,4$. Carrying out the same calculation for $\mathcal{B}_-$, one obtains
\begin{equation}
    \mathcal{B}_-(t_1,t_2,\sigma,\delta)\approx -c_1(\delta'+(A_1(\sigma-A_2)^2+A_3))\,,
\end{equation}
which justifies \eqref{eq:origin-near-BC}. The exact forms of $A_i$ in terms of $c_j$ are not particularly illuminating, but near $\delta t=\delta t_c$ we have
\begin{align}
    A_1&=-\left.\frac{c_2}{c_1}\right\vert_{\delta t_c}+O(\delta t-\delta t_c)\\
    A_2&=-\left.\frac{c_4\, \partial_{\delta t}c_0}{2c_1 c_2}\right\vert_{\delta t_c}(\delta t-\delta t_c)+O\qty((\delta t-\delta t_c)^2)\\
    A_3&=-\left.\frac{\partial_{\delta t}c_0}{c_1}\right\vert_{\delta t_c}(\delta t-\delta t_c)+O\qty((\delta t-\delta t_c)^2)\,.
\end{align}
One finds that $A_1(\delta t_c)<0$ and $\partial_{\delta t}A_3(\delta t_c)>0$, which supports the statements made around \eqref{eq:origin-near-BC}.

\bibliographystyle{utphys}
\bibliography{reference}

\end{document}